# Fast Calculation of Probabilistic Power Flow:
# A Model-based Deep Learning Approach

Yan Yang, Zhifang Yang, Juan Yu, Baosen Zhang, Youqiang Zhang, and Hongxin Yu

***Abstract*-- Probabilistic power flow (PPF) plays a critical role in power system analysis. However, the high computational burden makes it challenging for the practical implementation of PPF. This paper proposes a model-based deep learning approach to overcome the computational challenge. A deep neural network (DNN) is used to approximate the power flow calculation and is trained according to the physical power flow equations to improve its learning ability. The training process consists of several steps: 1) the branch flows are added into the objective function of the DNN as a penalty term, which improves the approximation accuracy of the DNN; 2) the gradients used in the back propagation process are simplified according to the physical characteristics of the transmission grid, which accelerates the training speed while maintaining effective guidance of the physical model; and 3) an improved initialization method for the DNN parameters is proposed to improve the convergence speed. The simulation results demonstrate the accuracy and efficiency of the proposed method in standard IEEE and utility benchmark systems.***

## I. INTRODUCTION

RENEWABLE energy resources have developed rapidly on a global scale in recent years [1], [2]. Consequently, the uncertainties in power systems have increased dramatically due to the integration of these intermittent resources. The surge of uncertainties has produced significant impacts on all sectors of power systems and makes the safe and stable operations of power grids challenging [1]. *Probabilistic power flow* (PPF) is an important tool to mitigate the impact of uncertainties by quantifying how the randomness of node injection power propagates to the bus voltage, power flow, and other system operating states. The knowledge of the uncertainties can inform the operator about various planning and operation problems [3]. Hence, for power systems with high-penetration renewables, there has been a recent push to implement PPF in practical operations. However, while the PPF has been extensively studied in academia for decades, there have been relatively few practical implementations of the PPF in power industries. The major reason is the heavy computational burden of the PPF method.

PPF analysis methods can be generally divided into analytical methods and numerical methods. The former attempts to directly work with the distributions of uncertainties, with approaches like cumulant methods [4], [5], point estimation methods [6], [7], Cornish-Fisher expansion [8], [9], and the generalized polynomial chaos method [10]. These analytical methods are considered to be computationally tractable because they directly integrate the probabilistic distribution of uncertainty factors (e.g., renewable energy source and load demand) into a few deterministic formulas. However, many practical distributions and systems do not lend themselves to these types of direct analysis. For example, in [4], [5], the power flow equations are assumed to be linear. In addition, the probability density function of output variables is approximated by their statistical moments/cumulants in [6]-[10]. These assumptions may not hold in real system of interest.

The second approach is to work with samples drawn from the distributions instead of the distributions themselves, which is the focus of this paper. The Monte-Carlo approach that can provide the accurate results is the representative of numerical approaches. This method typically contains two stages. In the first stage, a large number of samples are drawn randomly according to the probability distribution of net power injections. Then, the power flow calculation is performed for each sample and the statistics of the corresponding solutions (e.g., mean value, standard deviation and probability distribution of system states) are analyzed. Therefore, this approach involves repeatedly solving an enormous number of power flow problems. Although individual power flow calculation is not complex, the cumulative computational burden of so many calculations is considerable.

There are two ways to reduce the computational complexity of Monte-Carlo approach. The first is to reduce the number of samples by using a smaller yet still "representative" subset. A number of techniques have been developed, including importance sampling [11], Latin hypercube sampling [12], Latin supercube sampling [13], and Quasi-Monte sampling [14] and others. However, even if these improved sampling methods are used, a large number of samples may still be required to accurately reflect the uncertainty characteristics of the solutions. Parallel methods have also been proposed to improve the efficiency of PPF. The parallel methods can dramatically accelerate the speed using multiple GPUs [15] or cloud-computing platforms [16] without loss of accuracy.

Y. Yang, Z. Yang, and J. Yu are with Chongqing University, Chongqing 400044, China (e-mail: cquyangyan2012@163.com, yzf1992@cqu.edu.cn, yujuancqu@qq.com).

B. Zhang is with the Department of Electrical and Computer Engineering, University of Washington, Seattle, WA 98195, USA (zhangbao@uw.edu).

Y. Zhang, H. Yu are with State Grid Chongqing Electric Power Company Electric Power Research Institute, Chongqing 401123, China (e-mail: zyq113528@126.com, smileyapple@163.com).

However, the significant infrastructural investment of these methods limits their adoption in practice.

A comparatively less studied approach is to accelerate the power flow calculation of each sample. This approach can improve the computational efficiency of the PPF with any given number of samples, and thus it can be combined with the improved sampling techniques and parallel methods. For example, DC power flow can be used to convert nonlinear AC power flow into a linear program, and thus solved more efficiently. However, the DC power flow model cannot be applied when voltage magnitudes and reactive powers are of interest or when the accuracy of solutions is important [17].

This paper proposes to use the artificial neural networks to approximate and speed up power flow calculations [18], [19]. Since a PPF subjects a large number of samples to the same computational task (i.e., power flow calculation), it can be naturally formulated as a machine learning problem. Ref. [18] utilizes this idea and develop a control scheme via a *radial basis function* (RBF) neural network. In [19], RBF-based power flow is applied to the probabilistic PPF. However, the RBF is a shallow neural network, which does not always extract complex feature needed to approximate AC power flow equations. Besides, the RBF-based power flow calculations are cast as pure data-driven problems, where the goal is to learn the best neural network relating input to output without any guidance of the physical power flow model.

As shown by many successful applications in signal processing, *deep neural networks* (DNNs) have the ability to extract more abstract and complex features from data compared with shallow networks [20], [21]. Hence, DNNs show a promising way to approximate the power flow model and tackle the computational challenge of the PPF problem. The power flow calculation of the PPF can be regarded as a nonlinear function between the system operating condition (input) and the power flow solution (output).[1]

It turns out just making the neural network deeper is insufficient, since the underlying physical model is still not considered. In fact, we are dealing with the power flow equations, which are completely known. Therefore, we explicitly embed the physical model into the training process of the DNNs. This leads to both faster training and better test performances. To the best of our knowledge, similar methods have not been reported in current studies.

The main contribution of this paper is to construct a DNN to approximate the power flow calculation, which significantly improves the calculation efficiency of the PPF. The power flow model is used to improve the learning ability of the neural network, which has the following three features:

1. A composite objective function of the DNN is proposed based on the branch flow equations. The modified DNN more effectively extracts the nonlinear features of the power flow equations compared with a DNN trained via the standard least square loss on the output.

2. The training process is simplified to accelerate the training while remaining similar accuracy. Similar to the fast decoupled power flow, we remove the gradients from voltage magnitudes to the active and reactive branch powers and the gradients from phase angles to reactive powers in the learning process.

3. Both rectifier linear unit and linear activation function are used, and a new initialization method is designed to improve the training speed of DNNs.

Using the proposed method, DNNs can be trained faster with improved approximation accuracy compared with pure data-driven deep learning method. Ultimately, the calculation speed of the PPF can be improved by at least three orders of magnitude while maintaining similar accuracy compared with standard iterative power flow solvers.

## II. BASIC STRUCTURE OF THE PROPOSED METHOD

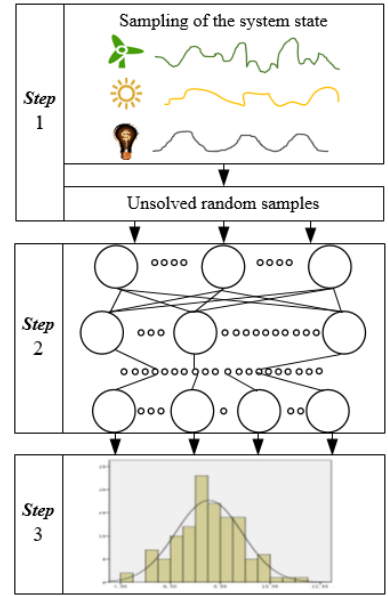

Fig. 1. Outline of a PPF based on a DNN. Step 1 draws samples of active and reactive power injections from the uncertainty distributions. In Step 2, a DNN translates these samples to complex voltages at each bus. Then statistical analysis is performed in Step 3.

The outline of a PPF based on a DNN is shown in Fig. 1. *Step* 1 is to sample the system states based on the probability distribution of net injections. In *Step* 2, the DNN directly computes the power flow solutions of all these samples. To be specific, the active and reactive power injections of uncertainty buses are treated as the input feature vector and the complex bus voltages are treated as the output feature vector. It dramatically accelerates the computation speed of the PPF problem since all the power flow operations are feed-forward function evaluations. *Step* 3 computes and analyzes the statistical indexes based on all of the samples, including the mean value, the standard deviation and the probability distributions of relevant events. In *Step* 1, any sampling algorithms can be used no matter a DNN or a power flow solver is used in *Step* 2. The DNN only needs to be trained once offline, and different operating conditions can be handled by this single network because of its generalization capability.

---

[1] The nonlinearity of DNNs may lead to the outputs of power flow with large errors, but we have observed that these outliers are extremely rare. Most samples of the power flow solutions are effectively learned, and the statistical indexes of interest (e.g., mean value, standard deviation and probability density function of system states) would be satisfactory.

Therefore, the crux of the DNN-based PPF is how well the DNN is able to approximate the power flow model. Hence, this paper focuses on the construction of the DNN and improving its performance via power flow model-based method and novel initialization.

## III. MODEL-BASED DEEP LEARNING TECHNIQUES

In this section, a modified loss function is described to guide the training process of DNNs, with the goal of improving the training speed and testing accuracy of DNNs by using the physics of power flow in the transmission grid.

### A. Feature vector selection

Essentially, the DNN mines the nonlinear features of the PPF by quantifying the effect of the input changes on the output. The PPF considers the effect of bus uncertainties on power flow results, therefore the input feature vector $X$ is designed as the injection power of all renewable energy sources and load demands.

The PPF requires repeatedly solving a tremendous number of deterministic power flow problems. The PPF output is concerned with bus voltages and branch flows. Therefore, the standard polar coordinate form of power flow problem is given by the following formulations [22], [23].

$$P_{ij} = G_{ij}\left(V_i^2 - V_i V_j \cos\theta_{ij}\right) - B_{ij} V_i V_j \sin\theta_{ij} \quad (1)$$

$$Q_{ij} = -B_{ij}\left(V_i^2 - V_i V_j \cos\theta_{ij}\right) - G_{ij} V_i V_j \sin\theta_{ij} \quad (2)$$

where $P_{ij}$ and $Q_{ij}$ are the active and reactive branch powers from the $i^{th}$ bus to the $j^{th}$ bus, respectively; $V_i$ is the voltage magnitude at bus $i$; $\theta_{ij}$ is the voltage phase angle difference between bus $i$ and bus $j$; $G_{ij}$ and $B_{ij}$ are the conductance and susceptance between the $i^{th}$ bus and the $j^{th}$ bus, respectively.

Because the vector of complex bus voltages (magnitudes and angles) completely describes the states of the system, we choose them as the output feature vector $Y$ of the neural network. Other quantities of interest, such as the branch flows, can be computed once the voltages are known. After the neural networks are trained, computing the output given an input can be done extremely efficiently since no iterations are required

### B. Loss Function Design Based on the Power Flow Equations

Essentially, the process of training a DNN is a fitting problem. Recall a DNN is a parameterized function, with the parameters conventionally denoted as $\theta = \{w, b\}$, where $w$ are the weights between the layers and $b$ are the biases. These parameters are optimized during training process to minimize the discrepancies between the outputs of the DNN and the labels. The squared difference is conventionally chosen as the loss function:

$$loss = \frac{1}{2m}\left\| Y_{out} - f_\theta^L\left(\cdots f_\theta^1\left(X_{in}\right)\right)\right\|^2, \quad (3)$$

$$f_\theta^i(X) = R_i\left(w_i X + b_i\right), \quad (4)$$

where $m$ is the number of training samples in each epoch; $L$ is the number of layers; $Y_{out}$ is the normalized output vector of the DNN; $X_{in}$ is the normalized input feature vector; $R_i$ is the activation function at the $i^{th}$ layer. The weight matrix $w_i$ is a

$n_{i+1} \times n_i$ matrix, and the biased vector $b_i$ is a $n_{i+1}$–dimensional vector, where $n_i$ is the number of neurons at the $i^{th}$ layer.

The loss function in (3) only induces the accuracy of the output feature vectors (bus voltage magnitudes and phase angles). Using this neural network, the accuracy of other quantities, such as the branch flows, may not be satisfactory because of the accumulation of errors and the nonlinear relationship between voltages and power flows. Therefore, we augment the loss function by explicitly adding branch flow equations as penalty terms into the objective. We can do this since once the voltages are known, we can explicitly compute the branch flows, and they act as side information to further enhance the learning efficiency. The modified loss function is

$$loss_{new} = loss + J\left(P_{out}, \hat{P}_{out}\right) + J\left(Q_{out}, \hat{Q}_{out}\right), \quad (5)$$

where

$$J\left(P_{out}, \hat{P}_{out}\right) = \frac{1}{2m}\left\| P_{out} - \hat{P}_{out}\right\|^2, \quad (6)$$

$$J\left(Q_{out}, \hat{Q}_{out}\right) = \frac{1}{2m}\left\| Q_{out} - \hat{Q}_{out}\right\|^2. \quad (7)$$

In (5), $P_{out}$ and $Q_{out}$ are $n_{brc} \times m$ matrices, which contain the normalized actual branch power flows; $\hat{P}_{out}$ and $\hat{Q}_{out}$ are the normalized values computed from the output of the DNN. We normalize the flows to make each component in (5) comparable and weight each component equally.

Because the learning target (i.e., the power flow model) is directly added to the objective function, the modified loss function will quickly guide the parameters $\theta = \{w, b\}$ toward an accurate approximation of power flow calculations. To illustrate the updating process, the weight matrix, $w$, is used as an example. The weights are updated using back-propagation algorithm (e.g., RMSProp [24]):

$$w_{(i,t+1)} = w_{(i,t)} - \Delta w_{(i,t)}, \quad (8)$$

$$\Delta w_{(i,t)} = \frac{\eta}{\sqrt{R_{(i,t)}} + \varepsilon} \odot dw_{(i,t)}, \quad (9)$$

$$R_{(i,t)} = \rho * R_{(i,t-1)} + \left(1-\rho\right) * dw_{(i,t)} \odot dw_{(i,t)}, \quad (10)$$

$$dw_{(i,t)} = \frac{1}{m}\sum_{k=r}^{r+m} \frac{\partial loss_{new}}{\partial w_{(i,t)}}, \quad (11)$$

where $w_{(i,t)}$ is the weight parameters from the $i^{th}$ layer to the $i+1^{th}$ layer at the $t^{th}$ parameters update; $r$ is the sequence number of initial samples in this batch; and $m$ is the sample size of this batch. In this paper, we use the following parameters: $\rho = 0.99$, $\varepsilon = 1 \times 10^{-8}$ and $\eta = 0.001$, which is a common setting in existing software.

The difference of the parameter updating between the loss functions in (3) and (5) shows up in (11). From (6) and (7), we can further decompose it into

$$dw_{(i,t)} = d\left(i\right)^t y_i / m, \quad (12)$$

where $y_i$ is the output of neurons at the $i^{th}$ layer, and

$$d\left(i\right) = d\left(i+1\right) w_{i,t-1} \odot \max\left(0, y_i\right), \quad (13)$$

$$d\left(L\right) = d_1 + d_2 + d_3, \quad (14)$$

$$d_1 = \hat{Y}_{out} - Y_{out}, \quad (15)$$

$$d_2 = \frac{(\hat{P} - P)}{std(P)} \odot \frac{\partial P}{\partial \hat{Y}_{out}} \times \frac{1}{std(P)} = \frac{(\hat{P} - P)}{std(P)} \odot \frac{\partial P}{\partial \hat{Y}} \times \frac{std(Y)}{std(P)}, (16)$$

$$d_3 = \frac{(\hat{Q} - Q)}{std(Q)} \odot \frac{\partial Q}{\partial \hat{Y}_{out}} \times \frac{1}{std(Q)} = \frac{(\hat{Q} - Q)}{std(Q)} \odot \frac{\partial Q}{\partial \hat{Y}} \times \frac{std(Y)}{std(Q)}, (17)$$

where $d(L)$ is the derivative of the loss function with respective to the DNN outputs. In equations (16) and (17), $Y$ is the output vector of the PPF; $\hat{Y}_{out}$ is the (normalized) output of the DNN; $\hat{Y}$ is the unnormalized value of $\hat{Y}_{out}$; and $\odot$ is the Hadamard (entry-wise) product.

From (12)-(14), we can see that $d(L)$ will directly affect the updating direction of the weight parameters $w$. The modified loss function guides the updating of the DNN parameters by modifying $d(L)$. The following content focuses on analyzing the impact of the modified function on the value of $d(L)$.

Using the conventional loss function (3), $d(L)$ only contains $d_1$. The added plenty terms (6) and (7) affect $d(L)$ via $d_2$ and $d_3$, respectively. From (8)-(17), the modified loss function in (5) increases the updating step size of the weight parameters $w$ when the updating direction simultaneously reduces (3), (6), and (7). Meanwhile, the proposed loss function is expected to reduce/prevent DNN over fitting the bus voltages when the parameter updating directions for (3), (6), and (7) are different. Numerical studies in Section V demonstrate the proposed method is effective to promote convergence in training.

We can also control the contribution of $d_2$ and $d_3$ to the output feature vector ($Y_{out} = [V, \theta]$) which contains the voltage magnitude and the phase angle. Therefore, two contribution weights $\alpha$ and $\beta$ for updating the voltage magnitudes and the phase angles are shown in below:

$$d_V[L] = d_{1,V} + \alpha(d_{2,V} + d_{3,V})$$
$$d_\theta[L] = d_{1,\theta} + \beta(d_{2,\theta} + d_{3,\theta}) \quad . \quad (18)$$
$$d[L] = [d_V[L], d_\theta[L]]$$

where $d_V[L]$, $d_{1,V}$, $d_{2,V}$, and $d_{3,V}$ are the partial derivative of function (5), (3), (6), (7) of the voltage magnitude (similarly for the $d_\theta[L]$, $d_{1,\theta}$, $d_{2,\theta}$, and $d_{3,\theta}$).

This strategy amounts to a form of regularization when training DNNs. An intuitive to set $\alpha$ and $\beta$ is that the contribution of $d_2 + d_3$ to the output feature vector should not be much larger than that of $d_1$. The following empirical formulas are used to determine the values of $\alpha$ and $\beta$.

$$\alpha = 0.5 \times \frac{max(abs(d_{1,V}))}{max(abs(d_{2,V} + d_{3,V}))}$$

$$\beta = 0.5 \times \frac{max(abs(d_{1,\theta}))}{max(abs(d_{2,\theta} + d_{3,\theta}))} \quad (19)$$

where $max$ is a function that returns the maximum value of a matrix, and $abs$ is the (entry-wise) absolute value function.

### C. Model-based Simplification of the Learning Process

The modified loss function can guide the training process and reduce the number of epochs needed to reach a satisfactory training error. However, it makes each epoch more computationally expensive. To mitigate this potentially adverse situation, we use the characteristics of the transmission grid to simplify the training process.

By standard calculations, the power flow sensitivities are:

$$\frac{\partial P_{ij}}{\partial \theta_i} = G_{ij} V_i V_j \sin\theta_{ij} - B_{ij} V_i V_j \cos\theta_{ij} = -\frac{\partial P_{ij}}{\partial \theta_j}, \quad (20)$$

$$\frac{\partial Q_{ij}}{\partial \theta_i} = -B_{ij} V_i V_j \sin\theta_{ij} - G_{ij} V_i V_j \cos\theta_{ij} = -\frac{\partial Q_{ij}}{\partial \theta_j}, \quad (21)$$

$$\frac{\partial P_{ij}}{\partial V_i} = 2G_{ij} V_i - V_j \left(G_{ij} \cos\theta_{ij} + B_{ij} \sin\theta_{ij}\right), \quad (22)$$

$$\frac{\partial P_{ij}}{\partial V_j} = -V_i \left(G_{ij} \cos\theta_{ij} + B_{ij} \sin\theta_{ij}\right), \quad (23)$$

$$\frac{\partial Q_{ij}}{\partial V_i} = -2B_{ij} V_i + V_j \left(B_{ij} \cos\theta_{ij} - G_{ij} \sin\theta_{ij}\right), \quad (24)$$

$$\frac{\partial Q_{ij}}{\partial V_j} = V_i \left(B_{ij} \cos\theta_{ij} - G_{ij} \sin\theta_{ij}\right). \quad (25)$$

The training process with the modified loss function can be activated by directly substituting equations (20)-(25) into (16) and (17). We find that some parts of (20)-(25) have little impact on the training process. Therefore, we can reduce the computational complexity through two simple steps.

#### i) Removing the guidance for voltage magnitudes

As mentioned above, the DNN extracts the nonlinear features/relationship of the PPF by quantifying the effect of the input changes on the output. In power systems, the voltage magnitude normally fluctuates within ±5%. However, the range of the voltage phase angle can reach more than 30 degrees. The change characteristic of the phase angle is much more complex than that of the voltage magnitude. Hence, the feature of the voltage magnitude is generally easier to be learned compared with the phase angle. The guidance of physical model is more necessary for the phase angle. Besides, because the standard deviation of the voltage magnitude is much less than that of the phase angle, it can be indicated in the (16) and (17) that the impact of model guidance on the voltage magnitude is much smaller than that of the phase angle.

From the perspective of computational complexity, the voltage magnitude incurs more computational cost than the phase angle. According to equations (14)-(17), the computation needs to execute all the equations in (20)-(25). If removing the guidance for the voltage magnitude, only (20) and (21) need to be executed. Therefore, the computational cost of the guidance for voltage magnitudes is approximately twice as that for the angle phases.

Consequently, according to the numerical analysis and computational complexity comparison, the guidance for the voltage magnitude is removed in the training process.

#### ii) Removing the guidance of reactive powers for phase angles

We further focus on the sensitivity of the phase angle according to formulations (20) and (21). In the transmission grid, we can generally have:

$$\left| B_{ij} \right| \Box \left| G_{ij} \right|,$$
$$B_{ij} > 0, G_{ij} < 0, and \ i \neq j \qquad (26)$$

Although the phase angle can change sharply with the operation condition, the phase angle difference of the two buses remains a relatively small number. Thus, we have

$$\sin \theta_{ij} < \cos \theta_{ij} . \qquad (27)$$

According to (26) and (27), it can be easily derived that the absolute value of formulation (21) is much less than formulation (20). Additionally, the absolute values of ($\hat{\boldsymbol{Q}}$-$\boldsymbol{Q}$) are generally less than those of ($\hat{\boldsymbol{P}}$-$\boldsymbol{P}$) because the voltage magnitude is easier to be learned than the phase angle. The standard deviation of the active branch power is generally less than that of the reactive branch powers because the active load demand is higher than the reactive load demand in most cases. Therefore, it can be concluded from (12)-(17) that the reactive branch power has a much lower impact on the training process compared to the active power flow. In consequence, we ignore the impact of reactive branch powers on the phase angle.

In summary, the equations used in the training process are simplified as follows.

$$(8) - (10) \qquad (28)$$

$$(12) - (13) \qquad (29)$$

$$\boldsymbol{d}(L) = \boldsymbol{d}_1 + \boldsymbol{d}_{1,\boldsymbol{\theta}} \qquad (30)$$

$$\boldsymbol{d}_1 = \hat{\boldsymbol{Y}}_{out} - \boldsymbol{Y}_{out} \qquad (31)$$

$$\boldsymbol{d}_{1,\boldsymbol{\theta}} = \boldsymbol{d}_{1,\boldsymbol{\theta}} + \alpha \frac{\left(\hat{\boldsymbol{P}} - \boldsymbol{P}\right)}{std(\boldsymbol{P})} \Box \frac{\partial \boldsymbol{P}}{\partial \hat{\boldsymbol{\theta}}} \times \frac{std(\boldsymbol{\theta})}{std(\boldsymbol{P})} \qquad (32)$$

## IV. ACTIVATION FUNCTION AND INITIALIZATION

### A. Designing Activation Function

The activation function has a crucial impact on the training process. *Rectifier linear unit* (ReLU) activation function has gained popularity in recent years compared with the more traditional sigmoid-like activation functions [25]. Additionally, a theoretical initialization method considering the ReLU function was derived in [26]. Therefore, we select ReLU as the activation function:

$$R_i(x) = \begin{cases} x & if \ x > 0 \\ 0 & if \ x \leq 0 \end{cases} . \qquad (33)$$

To improve the training efficiency of DNN, the input and output data of the PPF should be preprocessed to eliminate the numerical problem and the adverse influence of outlier samples in the training process. The z-score method shown in (34) is adopted to normalize the samples:

$$\boldsymbol{v}_{out} = \begin{cases} \dfrac{\boldsymbol{v} - v_{mean}}{v_{std}} & if \ v_{std} \neq 0 \\ \boldsymbol{v} - v_{mean} & if \ v_{std} = 0 \end{cases} , \qquad (34)$$

where $v_{mean}$ and $v_{std}$ are the mean and the standard deviation of vector $\boldsymbol{v}$, respectively. This method can effectively handle outliers, and only the mean and standard deviation of the historical statistics are required. Moreover, it preserves the shape of the input distribution better than other preprocessing methods such as the min-max method.

For the output, we cannot use ReLU activation function in the last layer of the neural network since it only outputs non-negative values. Therefore, the activation function of the last layer is designed to be linear:

$$R_L(x) = x . \qquad (35)$$

This idea has also been reported in many vision problems to give a wide range for the output [27].

### B. Initialization Method

The DNN parameter initialization can directly affect the training efficiency and even the convergence performance. In recent years, related studies can be categorized into two types: adding a pretraining stage to initialize the DNN [28], and random initialization approaches [26], [29]-[30]. The former requires more training time and may also lead the DNN to reach to a poor local optimum. Therefore, this paper focuses on the latter approach. The latest random initialization method that particularly adapts ReLU is proposed in [26].

In this paper, the ReLU and linear activation functions are both used. Hence, a new initialization approach is derived to improve the PPF learning efficiency. We define $\boldsymbol{y}_i$ and $\boldsymbol{z}_i$ as the activation vector and the argument vector at the $i$th layer.

#### i) Forward propagation case

We have:

$$\boldsymbol{z}_i = \boldsymbol{w}_i \boldsymbol{y}_i + \boldsymbol{b}_i, \ \ \boldsymbol{y}_i = R_i\left(\boldsymbol{z}_{i-1}\right) . \qquad (36)$$

Suppose that the elements of $\boldsymbol{y}_i$ and $\boldsymbol{z}_i$ are all independent to each other [26][29]. We then have:

$$Var\left[z_i\right] = n_i Var\left[w_i y_i\right], \qquad (37)$$

where $y_i$, $z_i$, and $w_i$ denote to the random variables of each element in $\boldsymbol{y}_i$, $\boldsymbol{z}_i$ and $\boldsymbol{w}_i$, respectively. We can obtain:

$$Var\left[z_i\right] = n_i Var\left[w_i\right] E\left(y_i^2\right) . \qquad (38)$$

As in [26] and [29], let $w_{i-1}$ have a symmetric distribution around zero and $b_{i-1} = 0$. Then, $z_{i-1}$ has a zero mean and has a symmetric distribution. Because $y_i$ is obtained by the ReLU function, we can have $E\left[y_i^2\right] = \frac{1}{2} E\left[z_{i-1}^2\right] = \frac{1}{2} Var\left[z_{i-1}\right]$. Then,

$$Var\left[z_i\right] = \frac{1}{2} n_i Var\left[w_i\right] Var\left[z_{i-1}\right] . \qquad (39)$$

Similarly, we have:

$$Var\left[z_{L-1}\right] = Var\left[z_1\right] \left( \prod_{i=2}^{L-1} \left( \frac{1}{2} n_i Var\left[w_i\right] \right) \right) . \qquad (40)$$

This equation is the key to the initialization design. A proper initialization method should avoid reducing or magnifying the magnitudes of input signals exponentially. Therefore, a sufficient condition is:

$$\frac{1}{2} n_i Var\left[w_i\right] = 1, i \geq 2 . \qquad (41)$$

When $i=1$, we have $z_1 = w_1 y_0$, where $y_0$ represents the input signal. Because the input signal has not gone through any activation function, we have

$$Var\left[w_1\right] = \frac{1}{n_1}. \tag{42}$$

In conclusion, the following equation can be obtained

$$Var\left[w_i\right] = \begin{cases} \dfrac{2}{n_i} & i \neq 1 \\[2mm] \dfrac{1}{n_i} & i = 1 \end{cases}, \quad i = 1,...,L-1. \tag{43}$$

*ii) Backward propagation case*

For the back-propagation process, we obtain the following:

$$\frac{\partial loss}{\partial \boldsymbol{y}_{i-1}} = \boldsymbol{w}_i^{\mathrm{T}} \frac{\partial loss}{\partial \boldsymbol{z}_i}, \quad \frac{\partial loss}{\partial \boldsymbol{z}_i} = R_i'(z_i) \frac{\partial loss}{\partial \boldsymbol{y}_i}, \tag{44}$$

where the superscript T denotes the transpose, and $\boldsymbol{w}_i^{\mathrm{T}}$ can be obtained by transposing the elements of $\boldsymbol{w}_i$. Similar to the forward propagation case, we assume that $w_i$ and $\partial loss/\partial z_i$ are independent of each other. We then have:

$$Var\left[\frac{\partial loss}{\partial y_{i-1}}\right] = n_{i+1} Var\left[w_i\right] E\left[\left(\frac{\partial loss}{\partial z_i}\right)^2\right]. \tag{45}$$

Additionally, $\partial loss/\partial y_{i-1}$ has a zero mean for all layers when $w_i$ is initialized by a symmetric distribution around zero. We assume that $R'$ and $\partial loss/\partial y_i$ are independent of each other. Because all but the last activation function are ReLU, we have:

$$E\left[\left(\frac{\partial loss}{\partial z_i}\right)^2\right] = \begin{cases} E\left[\left(\dfrac{\partial loss}{\partial y_i}\right)^2\right] = Var\left[\dfrac{\partial loss}{\partial y_i}\right] & i = L\text{-}1 \\[4mm] \dfrac{1}{2} E\left[\left(\dfrac{\partial loss}{\partial y_i}\right)^2\right] = \dfrac{1}{2} Var\left[\dfrac{\partial loss}{\partial y_i}\right] & i \neq L\text{-}1 \end{cases}. \tag{46}$$

If we consider a sufficient condition that the gradient is not exponentially large/small, we can then have:

$$Var\left[w_i\right] = \begin{cases} 1/n_{i+1} & i = L\text{-}1 \\[2mm] 2/n_{i+1} & i \neq L\text{-}1 \end{cases}, \quad i = 1,...,L-1. \tag{47}$$

Balancing the requirements in (43) and (47), we have:

$$Var\left[w_i\right] = \begin{cases} \dfrac{n_i + 2n_{i+1}}{n_i n_{i+1}} & i = L-1 \\[3mm] \dfrac{n_i + n_{i+1}}{n_i n_{i+1}} & i = 2,...,L-2 \\[3mm] \dfrac{2n_i + n_{i+1}}{n_i n_{i+1}} & i = 1 \end{cases}. \tag{48}$$

Consequently, the DNN weight parameter $w$ is initialized in a zero-mean Gaussian distribution whose standard deviation is shown in (49), and $\boldsymbol{b}$ is initialized as 0.

$$Std\left[w_i\right] = \begin{cases} \sqrt{\dfrac{n_i + 2n_{i+1}}{n_i n_{i+1}}} & i = L-1 \\[3mm] \sqrt{\dfrac{n_i + n_{i+1}}{n_i n_{i+1}}} & i = 2,...,L-2 \\[3mm] \sqrt{\dfrac{2n_i + n_{i+1}}{n_i n_{i+1}}} & i = 1 \end{cases}. \tag{49}$$

From formulations (43), (47) and (48), we can observe that the initialization of the middle-layer weight parameters $\boldsymbol{w}$ can meet all the criteria when the number of middle-layer neurons are equal. Therefore, we set the number of hidden neurons in each layer to be the same in our experiments.

## V. SIMULATION TESTS

The effectiveness of the proposed methods is verified using a modified IEEE 30-bus system, a modified IEEE 118-bus system and a 661-bus utility system.

### A. Test Information and Methods for Comparison

The penetrations of renewable energy sources are set at 30% in both the modified IEEE 30-bus and the 118-bus systems. The detailed description of the characteristics of the wind farms and photovoltaic stations can be found in [31]. The load demand is sampled randomly via a normal distribution with a standard deviation (10% of the mean).

Similar to [12]-[14], Monte-Carlo sampling method combined with Newton-Raphson algorithm (implemented in MATPOWER 6.0) is used as the PPF benchmark. Following methods are compared:

**M0**: A DNN only with ReLU activation function and the parameters are randomly initialized [26].

**M1**: A DNN with the designed activation function and the parameters are randomly initialized.

**M2**: A DNN with the designed activation function, the modified loss function and random initialization.

**M3**: A DNN with the designed activation function, the traditional loss function, and the proposed initialization method.

**M4**: A DNN with the proposed initialization method and the modified loss function.

**M5**: A DNN that is the same with M4, but the guidance for voltage magnitudes is removed in the training process.

**M6**: A DNN that is the same with M5, but the guidance of reactive powers for phase angles is removed.

These comparison methods and the corresponding intentions are listed in Table I.

TABLE I COMPARISON METHODS AND THE INTENTIONS

| Methods | Corresponding intention |
|---|---|
| **M1** and **M2** | Verify the effectiveness of the proposed loss function |
| **M0, M1** and **M3** | Verify the effectiveness of the proposed initialization method |
| **M1, M2, M3** and **M4** | Verify the effectiveness of the combination of the proposed initialization and the modified loss function. |
| **M1 M4,** and **M5** | Verify the effectiveness of removing the guidance for the voltage magnitude |
| **M1, M4, M5,** and **M6** | Verify the effectiveness of further removing the guidance of reactive branch power for the phase angle. |

Above methods have the same hyper parameters for each case, which are given in Table II. The number of validation samples and test samples is 10000. All the samples are generated according to the same distribution. The size of each batch is 100. The data source can be found in [32]. The training process is stopped if the DNN meets the condition of the early stop method [33] or the number of epochs reaches the threshold. All simulations are performed on a PC equipped with Intel(R) Core(TM) i7-7500U CPU @ 2.70GHz 32GB RAM. MATLAB is used as the simulation environment.

TABLE II HYPER PARAMETERS FOR DIFFERENT CASES

| Cases | Structure of DNN | Size of training dataset |
|---|---|---|
| Case 30 | [60 100 100 100 60] | 10000 |
| Case 118 | [236 200 200 200 236] | 20000 |
| Case 661 | [1322 500 500 500 500 500 1322] | 70000 |

To compare the performances of the different methods, the following indexes are used. $N_{epoch}$ refers to the number of epochs. $V_{loss}$ means the value of the original loss function (3). $P_{vm}$ refers to the proportion of the absolute error of the voltage magnitude that exceeds 0.0001 p.u. $P_{va}$ refers to the proportion of the absolute error of the phase exceeds 0.01 rad. $P_{pf}$ / $P_{qf}$ refers to the proportion of the absolute error of the active / reactive branch power that exceeds 5 MW. The solutions of PPF (e.g., mean value, standard deviation and probabilistic density function) are obtained by looking at the power flow results of all samples. Therefore, the indexes ($P_{vm}$, $P_{va}$, $P_{pf}$, and $P_{qf}$) are designed to quantify the accuracy of PPF by reflecting the power flow calculation accuracy of all test samples.

TABLE III PERFORMANCE COMPARISONS AMONG M1, M2, AND M3 FOR DIFFERENT CASES

| Cases | Method | $N_{epoch}$ | $V_{loss}$ | $P_{vm}$ | $P_{va}$ | $P_{pf}$ | $P_{qf}$ |
|---|---|---|---|---|---|---|---|
| Case 30 | M3 | 1000 | 1157 | 6.7% | 0.0% | 0.0% | 0.0% |
| | M2 | 1000 | 1175 | 4.0% | 0.0% | 0.0% | 0.0% |
| | M1 | 1000 | 1175 | 6.3% | 0.0% | 0.0% | 0.0% |
| Case 118 | M3 | 1000 | 44 | 2.7% | 0.0% | 3.5% | 0.0% |
| | M2 | 1000 | 75 | 4.7% | 0.1% | 7.6% | 0.3% |
| | M1 | 1000 | 80 | 5.4% | 0.0% | 8.1% | 0.4% |
| Case 661 | M3 | 500 | 3984 | 3.6% | 0.0% | 3.9% | 0.0% |
| | M2 | 500 | 4185 | 4.3% | 0.0% | 3.7% | 0.0% |
| | M1 | 500 | 4269 | 4.4% | 0.0% | 6.2% | 0.0% |

### B. Validations of the Initialization Method and Modified Loss Function

#### i ) Validation of the individual proposed method

Table III shows the performance comparisons among M1, M2 and M3 with the same number of epochs for different cases. It can be observed from Table III that the proposed loss function M2 can reduce the values of the original loss function M1 from 80, 4269 to 75, 4185 in Case 118 and Case 661, respectively. Besides, the values of $P_{vm}$ and $P_{pf}$ in Case 118 and Case 661 are both reduced while the values of $P_{va}$ and $P_{qf}$ are both close to 0.0%. In Case 30, although the value of $V_{loss}$ is not reduced by M2, the value of $P_{vm}$ still has an obvious improvement. Meanwhile, the values of $P_{va}$, $P_{pf}$, and $P_{qf}$ do not increase. Therefore, our proposed loss function M2 has higher accuracy than traditional loss function M1 for unseen samples (test samples). Therefore, the proposed idea to modify the loss function improves the generality ability of DNNs.

The necessity of the combination of ReLU activation function and linear function is verified firstly to demonstrate the effectiveness of the proposed initialization method. It can be observed from Fig. 2 that the DNN with our designed activation function M1 has an obvious priority than that only with ReLU activation M0 because the last linear activation function enables the DNN to capture a wider range of output.

On the basis of the designed activation function, it can also be observed from Table III that the proposed improved initialization method M3 can reduce $V_{loss}$ faster and obtain a better accuracy compared with the initialization method M1 in [26] in different cases. Specifically, the value of $V_{loss}$ can almost be cut in half by the proposed initialization method M2 in Case 118. Therefore, the derived initialization method can improve the convergence efficiency effectively.

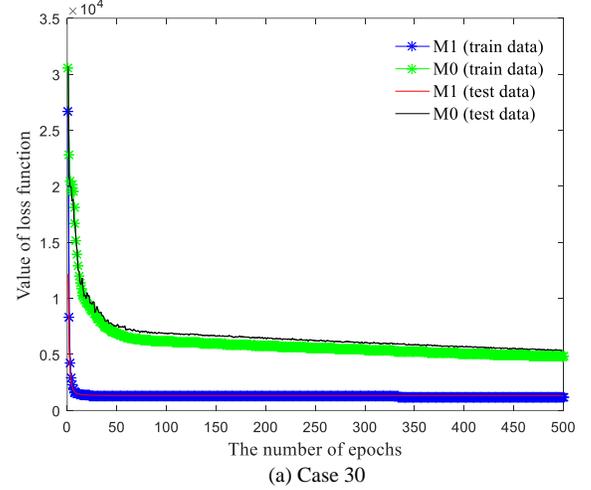

(a) Case 30

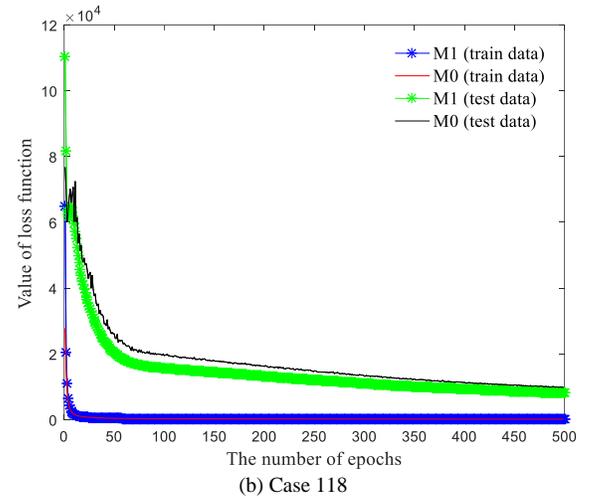

(b) Case 118

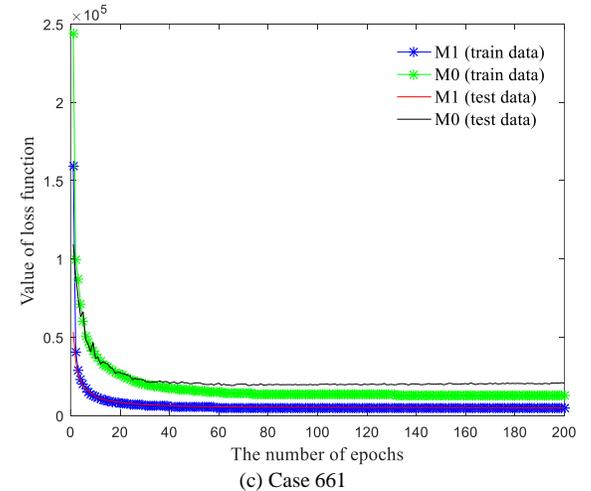

(c) Case 661

Fig. 2 Loss change process of training data and test data in different cases.

*ii ) Validation of the combination of proposed initialization method and modified loss function*

When the number of epochs is fixed, it can be seen from Table IV that the proposed method M4 can make all the indexes ($P_{vm}$, $P_{va}$, $P_{pf}$, and $P_{qf}$) meet the accuracy requirements (≤5%), while one or two indexes cannot achieve the requirements using method M1 (marked in bold). Additionally, comparing the results of Table IV and III, it can be concluded that the combination can achieve better results than the individual proposed method.

The training process is stopped as soon as all the indexes by DNN being no more than 5%, with the results shown in Table V. It can be observed that $N_{epoch}$ of M1 can be reduced by 68.7%, 71.7% and 61.3% by the proposed method M4 in the three cases, respectively. Additionally, it is worth noting that the values of all indexes can meet the accuracy requirements (≤5%) by M1 when the number of epochs reaches 375 in Table V. However, as the number of epochs increases to 500, the value of $P_{pf}$ by M1 has increased to 6.2% as shown in Table III, which means the DNN is over-fitting the data. Without the efficient guidance of the physical model and a suitable initial value, the branch flow accuracy cannot be guaranteed, even though the value of loss function has been reduced and most bus voltages are approximated quite well. Moreover, the value of $P_{pf}$ calculated by M4 is 3.4%. However, it is almost impossible to achieve an improvement of 1.6% from 5.0% using M1.

Summarizing the above discussions, the proposed loss function and initialization method can speed up the convergence dramatically and can reduce or prevent the DNN over fitting the bus voltages.

TABLE IV PERFORMANCE COMPARISONS BETWEEN M1 AND M4 UNDER THE SAME EPOCHS FOR EACH CASE

| Cases | Method | $N_{epoch}$ | $V_{loss}$ | $P_{vm}$ | $P_{va}$ | $P_{pf}$ | $P_{qf}$ |
|---|---|---|---|---|---|---|---|
| Case 30 | M4 | 1000 | 1175 | **3.6%** | 0.0% | 0.0% | 0.0% |
| | M1 | 1000 | 1175 | **6.3%** | 0.0% | 0.0% | 0.0% |
| Case 118 | M4 | 1000 | 41 | **2.8%** | 0.1% | **3.2%** | 0.0% |
| | M1 | 1000 | 80 | **5.4%** | 0.0% | **8.1%** | 0.4% |
| Case 661 | M4 | 500 | 3877 | 3.9% | 0.0% | **3.4%** | 0.0% |
| | M1 | 500 | 4269 | 4.4% | 0.0% | **6.2%** | 0.0% |

TABLE V PERFORMANCE COMPARISONS BETWEEN M1 AND M4 WHEN MEETING THE ACCURACY REQUIREMENTS

| Cases | Method | $N_{epoch}$ | $V_{loss}$ | $P_{vm}$ | $P_{va}$ | $P_{pf}$ | $P_{qf}$ |
|---|---|---|---|---|---|---|---|
| Case 30 | M4 | 507 | 1205 | ≤5% | ≤5% | ≤5% | ≤5% |
| | M1 | 1620 | 1146 | ≤5% | ≤5% | ≤5% | ≤5% |
| Case 118 | M4 | 569 | 64 | ≤5% | ≤5% | ≤5% | ≤5% |
| | M1 | 2012 | 41 | ≤5% | ≤5% | ≤5% | ≤5% |
| Case 661 | M4 | 145 | 4871 | ≤5% | ≤5% | ≤5% | ≤5% |
| | M1 | 375 | 4521 | ≤5% | ≤5% | ≤5% | ≤5% |

TABLE VI PERFORMANCE COMPARISONS AMONG M1, M5 AND M6 WHEN MEETING THE ACCURACY REQUIREMENTS

| Cases | Method | $N_{epoch}$ | $V_{loss}$ | $P_{vm}$ | $P_{va}$ | $P_{pf}$ | $P_{qf}$ |
|---|---|---|---|---|---|---|---|
| Case 30 | M6 | 775 ↑ | 1200 | ≤5% | ≤5% | ≤5% | ≤5% |
| | M5 | 576 ↑ | 1195 | ≤5% | ≤5% | ≤5% | ≤5% |
| | M1 | 1620 | 1146 | ≤5% | ≤5% | ≤5% | ≤5% |
| Case 118 | M6 | 622 ↓ | 62 | ≤5% | ≤5% | ≤5% | ≤5% |
| | M5 | 552 ↓ | 67 | ≤5% | ≤5% | ≤5% | ≤5% |
| | M1 | 2012 | 41 | ≤5% | ≤5% | ≤5% | ≤5% |
| Case 661 | M6 | 97 ↓ | 4887 | ≤5% | ≤5% | ≤5% | ≤5% |
| | M5 | 106 ↓ | 4887 | ≤5% | ≤5% | ≤5% | ≤5% |
| | M1 | 375 | 4521 | ≤5% | ≤5% | ≤5% | ≤5% |

## C. Validation of the Model-based simplification

### i) Performance comparison

Table VI shows the performance comparison among M1, M5, and M6. It can be observed that method M5 still has a significant advantage over M1. Comparing Table VI with Table V, it can be observed the number of epochs to meet the accuracy requirements using M5 is less than that of M4 in Case 118 and Case 661 (expressed by arrows). In Case 30, the number of epochs rises from 507 to 576, but it is tolerable in the small-scale case. Therefore, the simplification that ignoring the guidance of active branch powers and reactive branch powers for the voltage magnitude in the learning process can also maintain excellent performance, or even improve it.

The gradient between the reactive branch power and the phase angle is further ignored, and the numerical simulation of M6 is shown in Table VI. It can be observed that the decreasing or increasing trend of $N_{epoch}$ for M6 is the same as M5 compared with M4. In Case 661, $N_{epoch}$ of M6 is further reduced compared with M5, and $N_{epoch}$ of M1 can be reduced by 74.1%.

In conclusion, ignoring the guidance for the voltage magnitude and the guidance of reactive branch powers for the phase angle lead to similar or better performances, compared with the full information case.

### ii) Computation time comparison

Table VII shows the average computation time for each epoch with different methods. The total training time of above simulations can be calculated from this table. As expected, the simplified methods M5 and M6 cost less time than the basic model-based deep learning method M4 in all cases. In Case 30 and Case 118, the computation time with different methods does not have a significant difference. However, for a large practical power system, the difference is obvious significant.

With the removal of the guidance for the voltage magnitude, method M5 can reduce the computation time by 6.64 seconds for each epoch compared with M4. The computation time using M4 can be further reduced by 8.35 seconds by removing the guidance of reactive branch powers for phase angle M6. The speed advantage of the proposed model-based deep learning methods will increase with the number of epochs.

Above all, the proposed simplified model-based deep learning methods can reduce the computational pressure significantly compared with the basic model-based deep learning method M4, while maintaining comparable performance.

TABLE VII COMPUTATION TIME COMPARISON OF EACH EPOCH WITH DIFFERENT METHODS

| Cases | $t_{M1}(s)$ | $t_{M4}(s)$ | $t_{M5}(s)$ | $t_{M6}(s)$ |
|---|---|---|---|---|
| Case 30 | 0.13 | 0.26 | 0.21 | 0.20 |
| Case 118 | 0.68 | 1.64 | 1.25 | 1.09 |
| Case 661 | 35.21 | 51.53 | 45.07 | 43.18 |

## D. Performance of the Proposed Method for the PPF

The above numerical experiments have demonstrated the accuracy performance of the proposed method for calculating the PPF. The computation time comparison between the DNN and the benchmark PPF is shown in Table VIII. Note that the training time is not counted in the table because the DNN only needs to be trained once offline.

It can be seen that the proposed DNN method can accelerate the calculation speed by 1234 to 2040 times compared with the benchmark PPF. Therefore, the proposed approach provides an opportunity for the online application of PPF.

| Cases | Trained DNN (s) | Benchmark (s) | Acceleration ratio |
|-------|-----------------|---------------|--------------------|
| Case 30 | 0.04 | 81.59 | 2040 |
| Case 118 | 0.13 | 208.02 | 1600 |
| Case 661 | 0.82 | 1012.19 | 1234 |

## VI. CONCLUSION AND DISCUSSION

This paper proposes a model-based deep learning approach to quickly compute the solutions of power flow equations, with the main application of speeding up probabilistic power flow calculations. Based on the branch flow equations, a composite loss function is proposed to guide the training process. Combined with the physical characteristics of the transmission grid, the model-based training process method is simplified by removing the impact on voltage magnitudes and the relationship between reactive branch powers and phase angles. The proposed simplified method can accelerate the training speed while maintaining a comparable performance. In addition, an improved initialization method for the deep neural network parameters is derived to further improve the convergence efficiency. Simulation results using IEEE and utility test benchmarks demonstrate the effectiveness of the individually proposed methods against standard power flow solvers and other learning-based methods. We show the calculation speed of power flow problems can be accelerated by three orders of magnitude, thus allowing operators to consider 1000 times more samples under the same time constraints.

When the topology of the power grid or the distributions of renewable resources change, the trained DNN may not be applicable in these new scenarios. This challenge can be addressed by transfer learning techniques, where a trained neural network for a task is used as an initialization point to train a new neural network for other tasks [34]. Considering that the well-trained DNN has already extracted sufficient complex features for PPF, the transfer learning technique is a promising way to update the DNN parameters and reduce the effort to rebuild the DNN, and we plan to explore this direction in future works.